\documentclass{aa}
\usepackage{txfonts}
\usepackage[authoryear]{natbib}
\usepackage{natbib}
\usepackage{amssymb}
\bibpunct{(}{)}{;}{a}{}{,}
\usepackage{epsf}
\usepackage{graphicx}
\usepackage{epsfig}
\usepackage{graphics}
\usepackage{subfigure}
\usepackage[english]{babel}
\usepackage[figuresright]{rotating}
\usepackage{color}

\newcommand{\teff}{$T_{\rm{eff}}$}

\newcommand{\lL}{\ifmmode \log \frac{L}{L_{\sun}} \else $\log \frac{L}{L_{\sun
}}$\fi}
\newcommand{\mdot}{$\dot{M}$}

\newcommand{\vsini}{$V$~sin$i$}
\newcommand{\vinf}{$v_{\infty}$}

\newcommand{\kms}{km~s$^{-1}$}
\newcommand{\msun}{M$_{\sun}$}
\newcommand{\rsun}{R$_{\sun}$}
\newcommand{\zsun}{Z$_{\sun}$}

\begin{document}

\title{Evidence for quasi-chemically homogeneous evolution of massive stars up to solar metallicity}
\author{F. Martins\inst{1}
\and E. Depagne\inst{2}
\and D. Russeil\inst{3}
\and L. Mahy\inst{4}
}
\institute{LUPM--UMR 5299, CNRS \& Universit\'e Montpellier II, Place Eug\`ene Bataillon, F-34095, Montpellier Cedex 05, France \\
           \email{fabrice.martins@univ-montp2.fr}
\and
            Leibniz-Institut f$\rm{\ddot{u}}$r Astrophysik Potsdam, An der Sternwarte 16, 14482 Potsdam, Germany \\
\and
            LAM--UMR 6110, CNRS \& Universit\'e de Provence, rue Fr\'ed\'eric Joliot-Curie, F-13388, Marseille Cedex 13, France \\
\and
            Institut d'Astrophysique et de G\'eophysique, Universit\'e de Li\`ege, B\^at B5C, All\'ee du 6 Ao\^ut 17, B-4000, Li\`ege, Belgium \\
}

\date{Received / }

\abstract
{Long soft gamma ray bursts (LGRBs) are usually associated with the death of the most massive stars. A large amount of core angular momentum in the phases preceding the explosion is required to form LGRBs. A very high initial rotational velocity can provide this angular momentum. Such a velocity strongly influences the way the star evolves: it is chemically homogeneously mixed and evolves directly towards the blue part of the HR diagram from the main sequence.}
{We have shown that chemically homogeneous evolution takes place in the SMC, at low metallicity. We want to see if there is a metallicity threshold above which such an evolution does not exist.}
{We perform a spectroscopic analysis of H-rich early-type WN stars in the LMC and the Galaxy. We use the code CMFGEN to determine the fundamental properties (\teff, L) and the surface composition of the target stars. We then place the stars in the HR diagram and determine their evolution.}
{We show that both the LMC and Galactic WNh stars we selected cannot be explained by standard stellar evolution. They are located on the left of the main sequence but show surface abundances typical of CN equilibrium. In addition, they still contain a large amount of hydrogen. They are thus core-H burning objects. Their properties are consistent with chemically homogeneous evolution. We determine the metallicity of the Galactic stars from their position and Galactic metallicity gradients, and conclude that they have 0.6$<$Z$<$1.0. A moderate coupling between the core and the envelope is required to explain that stellar winds do not extract to much angular momentum to prevent a blueward evolution.}
{We have shown that chemically homogeneous evolution takes place in environments with metallicity up to solar. In view of the findings that some long gamma ray bursts appear in (super-)solar environments, such an evolution may be a viable way to form them over a wide range of metallicities. }


\authorrunning{Martins et al.}
\titlerunning{Quasi homogeneous evolution of WN stars}

\maketitle

%%%%%%%%%%%%%%%%%%%%%%%%%%%%%%%%%%%%%%%%%%%%%%%%%%%%%%%%%%%%%%%%%%%%%%%%%%%%%%%%%%%%%%%%%%%%%%%%%%%%%%%%%%%%%%%%%%%%%%%%%%%%%%%
%%%%%%%%%%%%%%%%%%%%%%%%%%%%%%%%%%%%%%%%%%%%%%%%%%%%%%%%%%%%%%%%%%%%%%%%%%%%%%%%%%%%%%%%%%%%%%%%%%%%%%%%%%%%%%%%%%%%%%%%%%%%%%%
\section{Introduction}
\label{s_intro}

Long soft Gamma-Ray Bursts (LGRB) are classically associated with the death of massive stars. The most convincing evidence for this is the clear association of some LGRBs with type Ib/c supernovae \citep{galama98,hjorth03}. They are also observed in star forming galaxies, often directly in sites where stars are currently being formed \citep[but see][]{hammer06}. The conditions required to produce LGRBs impose that its progenitor must have an envelope free of hydrogen so the jet can travel through it without being damped. The progenitor core must also retain a large angular momentum in order to produce a disk-jet structure around the black hole resulting from the core collapse. The collapsar model proposed by \citet{ww93} fulfills these conditions and is presently the main scenario invoked to explain the existence of LGRBs. 

The conditions required by the collapsar scenario, especially the large angular momentum of the core before explosion, have implications for the properties of the progenitor star. Since massive stars lose mass at a large rate due to their strong radiatively driven stellar winds \citep{cak,puls08}, they also lose a significant amount of angular momentum during their life. But massive stars winds also depend on metallicity \citep[\mdot$\propto Z^{0.6-0.8}$][]{mokiem07,vink01}: the lower the metal content, the smaller the radiative acceleration through lines and thus the lower the mass loss rate. The observation of larger rotational velocities in the LMC/SMC compared to the Galaxy \citep{martayan06,mokiem06} is an indirect confirmation of this effect (although the initial distribution of rotational velocities might be different too). Massive stars in low-Z environments are thus more likely to become LGRBs in the scenario where the angular momentum evolution is governed by mass loss (see \citet{petrovic05b} for alternative possibilities).. But an unusually large initial rotational velocity is also required since not all low-Z massive stars retain enough angular momentum at the end of their life to become a LGRB. 

Rapid rotation affects stellar evolution. \citet{maeder87} \citep[see also][]{langer92} showed that for very fast rotating massive stars, the mixing timescale becomes shorter than the nuclear timescale. In that case, the material produced in the stellar core is immediately mixed with the outer layers. The star is thus (quasi) chemically homogeneous. The larger fraction of helium in the outer layers reduces the opacity, causing the star to become hotter. Consequently, chemically homogeneous stars evolve directly to the blue part of the HR diagram from the main sequence.

We showed that quasi-chemically homogeneous evolution (CHE) was most likely followed by the star WR~1 in the SMC \citep{wnh09}. The analysis of its UV-optical-IR spectrum indicated a hot temperature (65000 K), placing the star on the blue side of the main sequence in the HR diagram. At the same time, we determined that WR~1 still contained a large amount of hydrogen in its envelope. This was consistent with its spectral classification (WN5h) typical of WN stars showing the presence of hydrogen lines in their atmosphere. Classical evolutionary tracks are not able to reproduce both the position in the HR diagram and the large hydrogen content. When they reach the position of WR~1, they have long burnt all of their hydrogen. Only chemically homogeneous evolution can explain the properties of WR~1. This is strengthened by our determination of the CNO content, fully consistent with the values expected from CNO equilibrium. SMC WR~1 is thus very likely a massive star in the core-H burning phase and evolving homogeneously towards the blue part of the HR diagram. It validates the concept that some stars follow CHE even if this does not necessarily mean that they will end as a LGRB. 

This study was focused on stars in the SMC. One may wonder whether CHE also happens at higher metallicity \citep[see also][]{graf11}. This is important in the context of the collapsar model. The first determinations of the metallicity of LGRB hosts tended to show that low-Z environments were preferred. For instance, \citet{modjaz08} showed that type Ic supernovae associated with LGRBs were located in galaxies with metallicities systematically lower than type Ic SN without GRB. Various analysis of LGRB hosts confirmed that low metallicity was preferred \citep{thone08,levesque11}. On the theoretical side, \citet{yl06} showed that stellar evolution at metallicities above $\sim$0.6 ended with too low angular momentum in the core to produce LGRBs. A threshold in metallicity around 0.6 \zsun\ for the formation of LGRBs seemed to exist. 

Things have changed recently on the observational side with the discovery of at least two LGRBs at super-solar metallicity \citep{graham09,levesque10}. In addition, the determination of the mass-metallicity relation for LGRB hosts indicates that solar and super-solar metallicities are not excluded, although the relation is shifted towards lower Z compared to the classical M-Z relation \citep{mannucci11}. The tendency of LGRBs to be located in actively star-forming galaxies explains this offset: such galaxies are currently mainly low-mass galaxies.

The question of the occurrence of CHE at high metallicity then arises. If massive stars never evolve homogeneously at high Z, this might favour alternative evolutionary paths to form LGRBs in high metallicity environments. \citet{petrovic05b} \citep[see also][]{gv12} showed that under the assumption of a weak coupling between the core and the envelope, the accreting star of a close binary, or a single star, might retain enough angular momentum after the red supergiant phase to produce a LGRB. 
In the present paper, we report on the analysis of hydrogen-rich WN stars in the LMC and the Galaxy. In a manner similar to the WNh stars in the SMC, we determine their physical properties and conclude about their evolution. We show that CHE is likely to exist in the LMC and the Galaxy, at solar metallicity. 

The paper is organized as follows: in Sect.\ \ref{s_obs} we describe our sample and the observations; we then explain how we determined the fundamental properties of the sample stars in Sect.\ \ref{s_models}; we then present our results in Sect.\ \ref{s_results} and discuss them in Sect.\ \ref{s_disc}. The conclusions are gathered in Sect.\ \ref{s_conc}.

%%%%%%%%%%%%%%%%%%%%%%%%%%%%%%%%%%%%%%%%%%%%%%%%%%%%%%%%%%%%%%%%%%%%%%%%%%%%%%%%%%%%%%%%%%%%%%%%%%%%%%%%%%%%%%%%%%%%%%%%%%%%%%%
%%%%%%%%%%%%%%%%%%%%%%%%%%%%%%%%%%%%%%%%%%%%%%%%%%%%%%%%%%%%%%%%%%%%%%%%%%%%%%%%%%%%%%%%%%%%%%%%%%%%%%%%%%%%%%%%%%%%%%%%%%%%%%%
\section{Sample stars and observations}
\label{s_obs}

\begin{table*}[]
\begin{center}
\caption{Sample stars and observational information} \label{tab_data}
\begin{tabular}{lrrrrrrr}
\hline
Star &  Optical spectrum  &  Date of optical observation & Resolution & & UV spectrum & Date of UV observation & Resolution \\
\hline
 Galaxy \\
\hline
WR7   &  VLT/UVES        & 26-27 Jan 2008                & 30000      & & IUE sp03541 & 6 Dec 1978             & 250 \\
WR10  &  VLT/UVES        & 26-27 Jan 2008                & 30000      & & IUE sp10748/sp13912 & 4 Dec 1980 / 6 May 1981      & 250 \\
WR18  &  VLT/UVES        & 26-27 Jan 2008                & 30000      & & IUE sp0933  & 18 May 1980      & 250 \\
WR128 &  NTT/FEROS       & 3 May 2012                    & 48000      & & IUE sp15101  & 26 Sep 1981      & 18000 \\
\hline
 LMC & \\
\hline
Bat 18 & SAAO/GS$^1$     & 28 Dec 1999                   & 900        & & IUE sp38236/sp38247 & 24 Fev 1990 & 250 \\
       &                 & 16-20 Jan 2001                & \\
       &                 & 1-7 Jan 2002                  & \\
Bat 63 & SAAO/GS$^1$     & 28 Dec 1999                   & 900        & & IUE sp09157/sp09158 & 31 May 1980 & 250 \\
       &                 & 16-20 Jan 2001                & \\
       &                 & 1-7 Jan 2002                  & \\ 
\hline
\end{tabular}
\tablefoot{1. SAAO/GS stands for SAAO observatory / Grating spectrograph. From \citet{foel03} }
\end{center}
\end{table*}

The optical data have been obtained from a variety of telescopes/instruments. Data for the LMC stars have been collected at SAAO and are those of \citet{foel03}. Details on the data reduction can be found in \citet{foel03}. We have used the average of all optical spectra in our analysis, to provide average stellar parameters. The optical spectrum of WR~128 was collected by one of us (LM) during run 089.D-0730(A)) on the 2.2m telescope equipped with FEROS spectrograph at the La Silla ESO observatory. The reduction procedure is identical to that described in \citet{san06} and was done under the MIDAS environment. Data for WR~7, WR~18 and WR~128 were obtained with the spectrograph UVES on ESO/VLT (run 080.D-0137(A)). 
UVES spectra were reduced using the UVES pipeline provided by ESO, which performs bias and inter-order background subtraction (object and flat-field), optimal extraction of the object (above sky, rejecting cosmic ray hits), division by a flat-field frame extracted with the same weighted profile as the object, wavelength calibration and rebinning to a constant wavelength and step, and merging of all overlapping orders.
In addition to the optical spectra, we have collected UV data from the IUE archive. Information on the spectra are given in Table \ref{tab_data}. When several IUE spectra are listed, we used the average spectrum to perform our spectroscopic analysis.

\begin{table*}[h]
\begin{center}
\caption{Photometry and distance of the target stars.} \label{tab_phot}
\begin{tabular}{llrrrrrrr}
\hline
Star & ST               &  U  &  B  &  V  &  J  &  H  &  K  &  distance \\ 
     &                  &     &     &     &     &     &     &  [kpc] \\
\hline
 Galaxy \\
\hline
WR7   & WN4             & 11.21 & 11.68 & 11.40 & 9.97 & 9.67 & 9.27 & 3.67\\
WR10  & WN5h            & 10.62 & 11.30 & 10.90 & 10.05 & 9.89 & 9.61 & 4.68\\
WR18  & WN4             & 11.10 & 11.40 & 10.60 & 8.57 & 8.21 & 7.68 & 2.20\\
WR128 & WN4(h)          & 9.59 & 10.54 & 10.46 & 9.97 & 9.84 & 9.62 & 9.37\\
\hline
 LMC & \\
\hline
Bat 18 & WN3h           & 13.57 & 14.45 & 14.67 & 14.63 & 14.49 & 14.40 & 50.12 \\
Bat 63 & WN4ha          & 13.62 & 14.45 & 14.60 & 14.84 & 14.62 & 14.50 & 50.12 \\
\hline
\end{tabular}
\tablefoot{References for the distances are: \citet{massey01} for WR10, \citet{vdh01} for WR18, 7, 128 . The distance modulus to the LMC is assumed to be 18.5.}
\end{center}
\end{table*}

%%%%%%%%%%%%%%%%%%%%%%%%%%%%%%%%%%%%%%%%%%%%%%%%%%%%%%%%%%%%%%%%%%%%%%%%%%%%%%%%%%%%%%%%%%%%%%%%%%%%%%%%%%%%%%%%%%%%%%%%%%%%%%%
%%%%%%%%%%%%%%%%%%%%%%%%%%%%%%%%%%%%%%%%%%%%%%%%%%%%%%%%%%%%%%%%%%%%%%%%%%%%%%%%%%%%%%%%%%%%%%%%%%%%%%%%%%%%%%%%%%%%%%%%%%%%%%%
\section{Modelling}
\label{s_models}

We have used the code CMFGEN \citep{hm98} to determine the stellar and wind parameters of the sample stars. CMFGEN solves the radiative transfer in the comoving frame in non-LTE conditions. A spherical geometry is adopted to correctly take the wind extension into account. The rate equations and the radiative equation are solved simultaneously through an iterative scheme until convergence of the level populations has been achieved. The temperature structure is set by the radiative equilibrium constraint. The velocity law (and consequently the density law through the mass conservation equation) is the result of the combination of an inner structure connected to a $\beta$ velocity law in the outer atmosphere. The inner structure is iterated: once the level populations have converged, the total radiative acceleration is calculated and the momentum equation is solved to provide a new velocity law. This solution is connected to the $\beta$ velocity law and the resulting velocity structure is used to compute a new atmosphere model with new level populations. A couple of these hydrodynamical simulations is performed. In the model computation, a constant microturbulent velocity of 50 \kms\ is adopted. The elements included in our models are H, He, C, N, O, Ne, P, Si, S, Fe, Ni. We use the solar abundances of \citet{ga07} as reference.

Once the atmosphere model is completed, a formal solution of the radiative transfer equation is performed to yield the emergent spectrum. A depth variable microturbulence velocity, from 10 \kms\ at the bottom of the wind to 10\% of the terminal velocity in the outter layers is adopted. The emergent spectrum is subsequently compared to the observed data to determine the stellar and wind parameters. In practice, we proceed as follows:

\begin{itemize}

\item[$\bullet$] \textit{Effective temperature}: we use the ionization balance method to constrain \teff. The ratio of \ion{N}{iv} to \ion{N}{v} lines is a first indicator. \ion{N}{iv} 4058, \ion{N}{iv} 5205, \ion{N}{iv} 5736, \ion{N}{iv} 7103, \ion{N}{v} 4519-4523, \ion{N}{v} 4944 are the main indicators. We also use the strength of the \ion{He}{ii} lines and, when observed, of \ion{He}{i} 5876 as a secondary temperature indicators. The \ion{Fe}{v} 1350-1400 forest in the UV range provides an additional constraint. For Wolf-Rayet stars, a meaningful quantity is the temperature at an optical depth of about 20 (T$*$). It is appropriate for comparison with temperatures provided by evolutionary models since it corresponds to the quasi-hydrostatic part of the atmosphere.

\item[$\bullet$] \textit{Luminosity}: the luminosity is derived from the fit of the spectral energy distribution (SED). Flux-calibrated IUE spectra as well as UBVJHK photometry are used to constraint \lL\ and the amount of extinction. The distances we adopted are listed in Table \ref{tab_phot}. For the Galactic objects, they rely on the correlation between the position of the target stars and OB associations. Due to uncertainties in these correlations and in the distance of Galactic associations, the errors on the luminosity of the Galactic WNh stars might be underestimated.

\item[$\bullet$] \textit{Mass loss rate}: it is derived from the absolute strength of the emission lines. H$\alpha$, H$\beta$, H$\gamma$ and \ion{He}{ii} 4686 are the main indicators. The UV resonance lines (\ion{N}{v} 1240, \ion{C}{iv} 1550, \ion{He}{ii} 1640, \ion{N}{iv} 1720) are also used. Generally the optical and UV wind-sensitive lines give consistent results. 

\item[$\bullet$] \textit{Terminal velocity}: The blueward extent of the UV resonance lines is the main indicator. The width of optical lines is the secondary diagnostic. 

\item[$\bullet$] \textit{Surface abundances}: The He/H ratio is constrained from the relative strength of H and HeII lines. This is different from the determination of the mass loss rate in which the absolute level of emission is used. The blends \ion{He}{ii}/H$\gamma$ and \ion{He}{ii}/H$\beta$ together with \ion{He}{ii} 4542 and \ion{He}{ii} 5412 are the main indicators. Carbon abundances are determined from \ion{C}{iv} 5802-5812. Nitrogen abundances result from the analysis of \ion{N}{iv} 4058, \ion{N}{v} 4519-23, \ion{N}{v} 4944, \ion{N}{iv} 5205, \ion{N}{iv} 5735 and, when available, \ion{N}{iv} 7103. 

\item[$\bullet$] \textit{Clumping factor}: we used mainly the red wing of the strong optical emission lines to determine the volume filling factor $f$ (in the micro-clumping formalism). According to \citet{hil91}, the electron scattering wing depends on the degree of inhomogeneity, being stronger in homogeneous winds. In practice, \ion{He}{ii} 4686 provides the best constraints. 

\item[$\bullet$] \textit{Slope of the velocity field}: we have used a value of 1.0 for the so-called $\beta$ parameter in our modelling of the Galactic targets. It provided a good fit of the Balmer and \ion{He}{ii} emission lines. For the LMC stars, we found that $\beta$=2.0 (respectively 1.5) lead to a better shape of the optical emission lines of Bat~18 (respectively Bat~63). 

\end{itemize}

The typical uncertainties have been estimated by varying the main parameters around the preferred values. When a clear deviation is observed in the fit, we assume we are seeing the effect of a one sigma deviation from the best set of parameters. The uncertainties are the following: 3000 K on the temperature; 0.10-0.15 dex on the luminosity; 0.2 dex on the mass loss rates; 200 \kms on the terminal velocity; 30 to 40\% on the surface abundances.

%%%%%%%%%%%%%%%%%%%%%%%%%%%%%%%%%%%%%%%%%%%%%%%%%%%%%%%%%%%%%%%%%%%%%%%%%%%%%%%%%%%%%%%%%%%%%%%%%%%%%%%%%%%%%%%%%%%%%%%%%%%%%%%
%%%%%%%%%%%%%%%%%%%%%%%%%%%%%%%%%%%%%%%%%%%%%%%%%%%%%%%%%%%%%%%%%%%%%%%%%%%%%%%%%%%%%%%%%%%%%%%%%%%%%%%%%%%%%%%%%%%%%%%%%%%%%%%
\section{Results}
\label{s_results}

%%%%%%%%%%%%%%%%%%%%%%%%%%%%%%%%%%%%%%%%%%%%
\subsection{Stellar parameters}

\begin{figure}[]
\centering
\includegraphics[width=9cm]{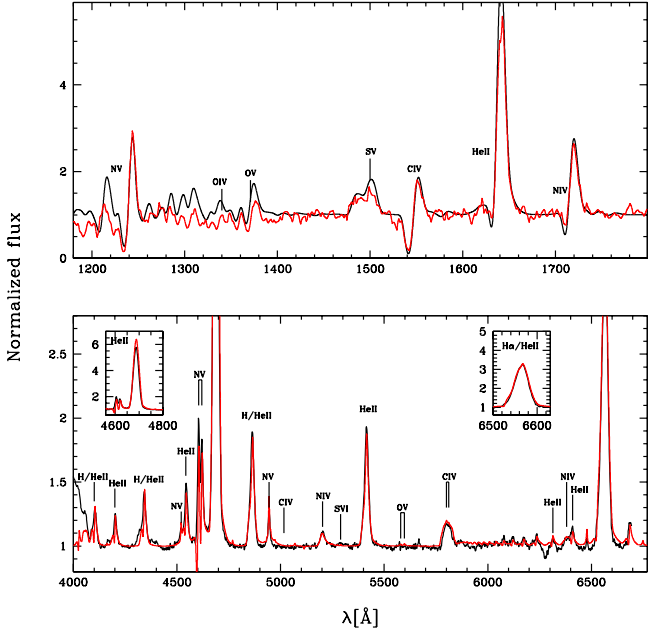}
\caption{CMFGEN model (red) compared to the observed spectrum (black) of star BAT~18.}
\label{fit_bat18}
\end{figure}

\begin{figure}[]
\centering
\includegraphics[width=9cm]{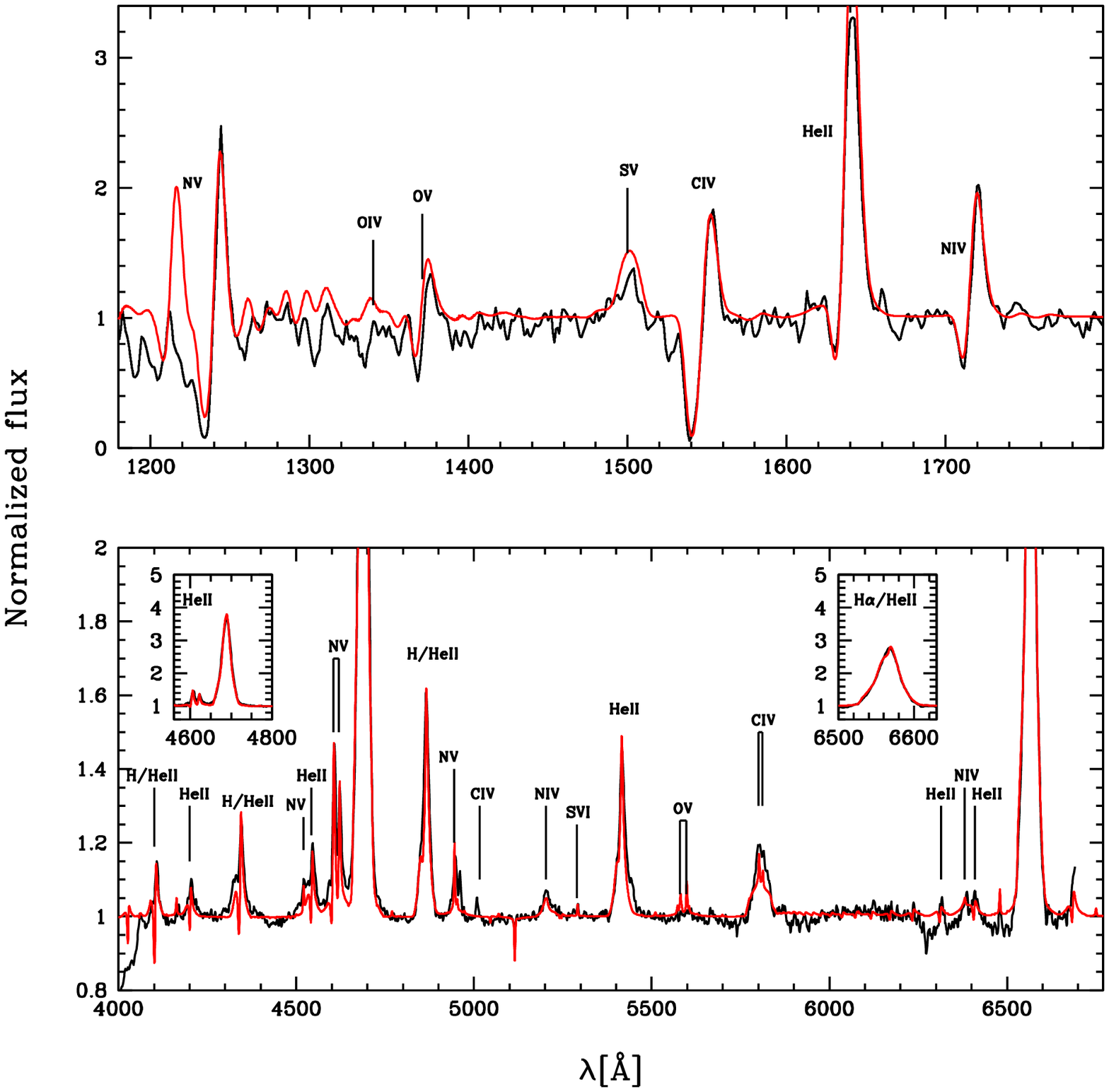}
\caption{CMFGEN model (red) compared to the observed spectrum (black) of star BAT~63.}
\label{fit_bat63}
\end{figure}

\begin{figure}[]
\centering
\includegraphics[width=9cm]{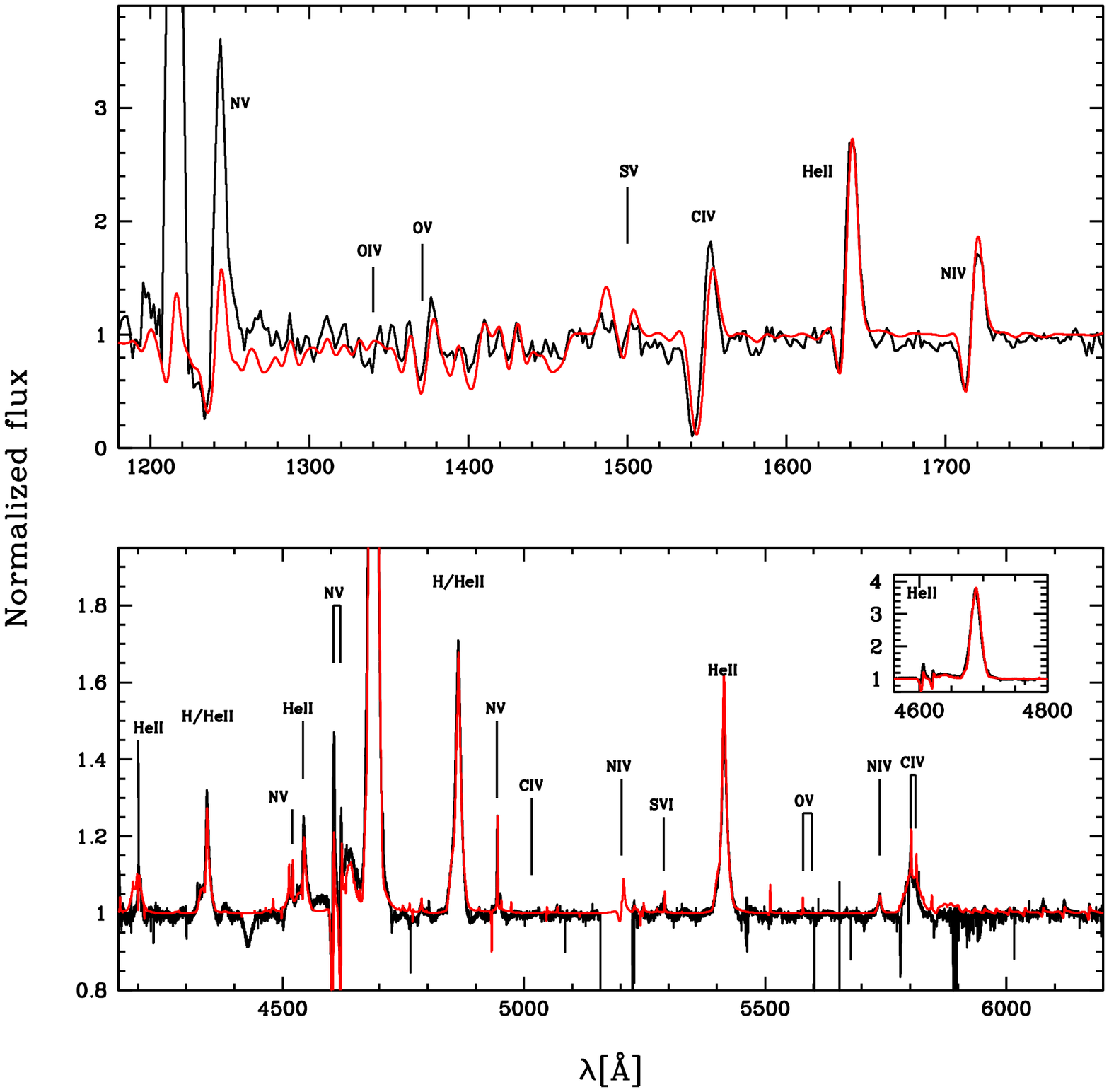}
\caption{CMFGEN model (red) compared to the observed spectrum (black) of star WR~10.}
\label{fit_wr10}
\end{figure}

\begin{figure}[]
\centering
\includegraphics[width=9cm]{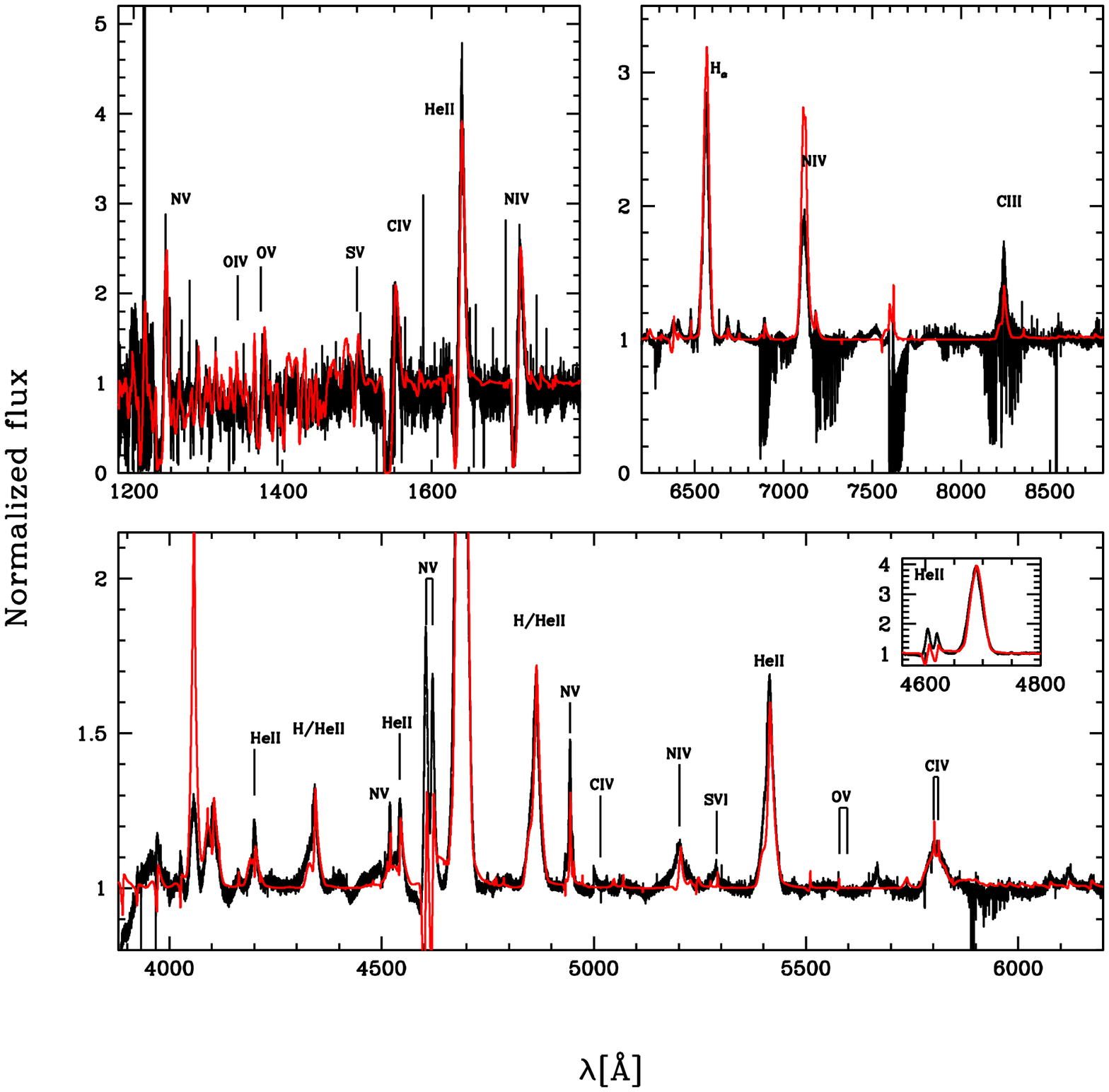}
\caption{CMFGEN model (red) compared to the observed spectrum (black) of star WR~128.}
\label{fit_wr128}
\end{figure}

\begin{figure}[]
\centering
\includegraphics[width=9cm]{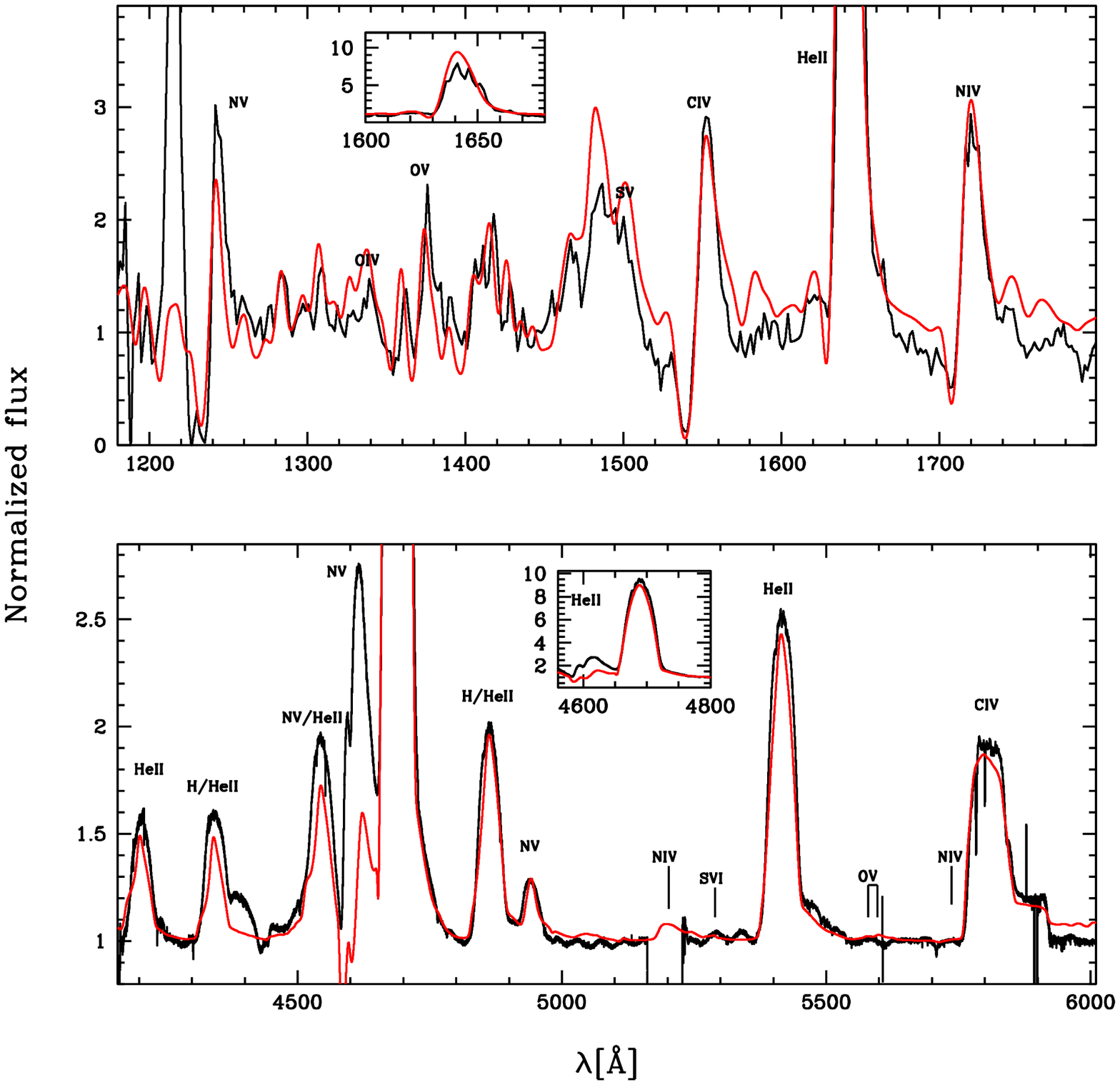}
\caption{CMFGEN model (red) compared to the observed spectrum (black) of star WR~18.}
\label{fit_wr18}
\end{figure}

\begin{figure}[]
\centering
\includegraphics[width=9cm]{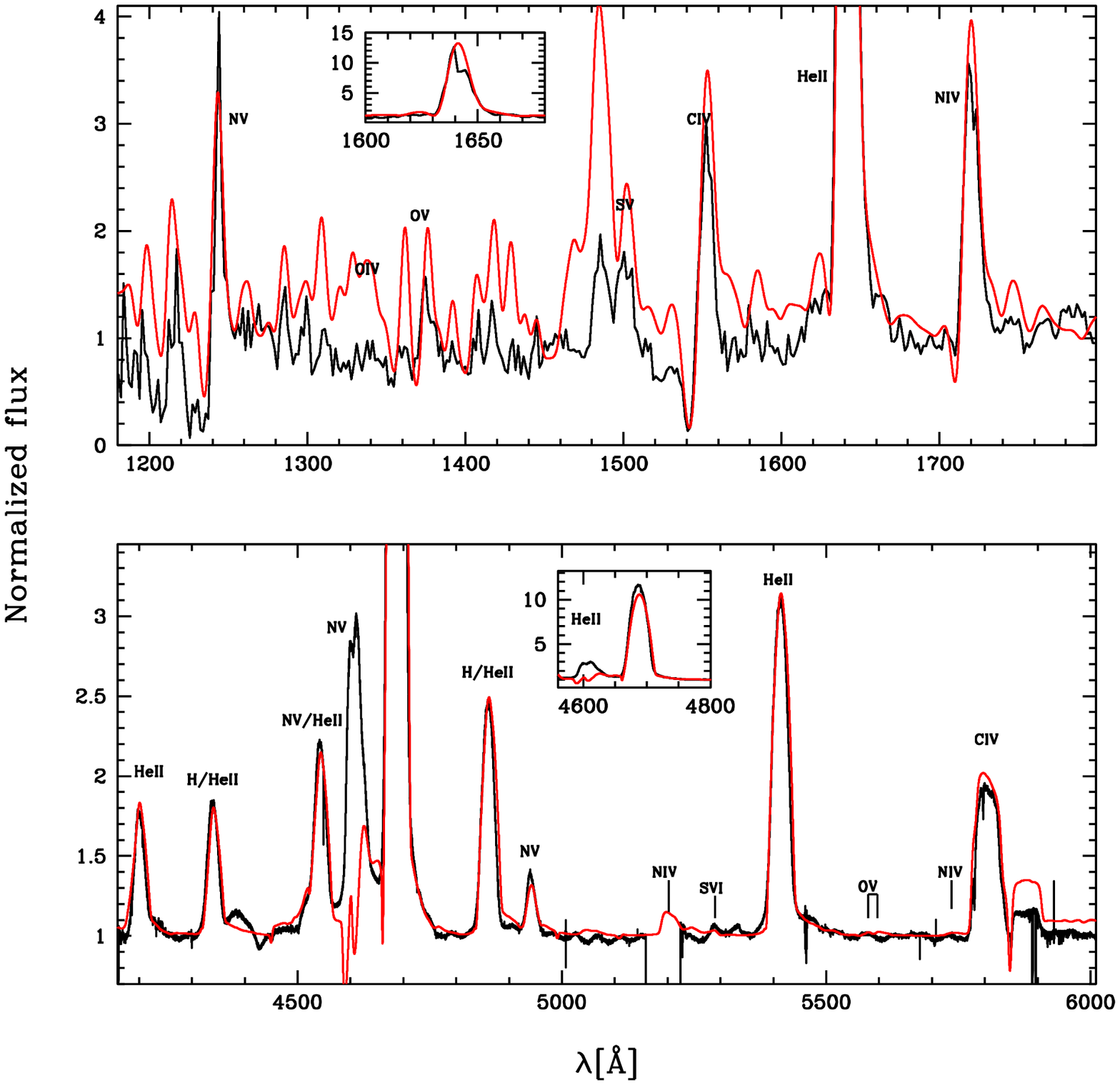}
\caption{CMFGEN model (red) compared to the observed spectrum (black) of star WR~7.}
\label{fit_wr7}
\end{figure}

The best fit to the observed spectra of our sample stars are shown in Figs.\ \ref{fit_bat18} to \ref{fit_wr7}. They are usually of excellent quality. Most lines are well reproduced. We note a difficulty to reproduce the \ion{N}{v} 4605-4620 features in Galactic stars. They are usually too weak compared to the observed spectrum. The other nitrogen lines are well reproduced, especially those used to determine the effective temperature. The reason for this discrepancy is not known at present. Similar problems were noted by \citet{bouret12}. The derived parameters are gathered in Table \ref{tab_res}.

\begin{figure}[]
\centering
\includegraphics[width=9cm]{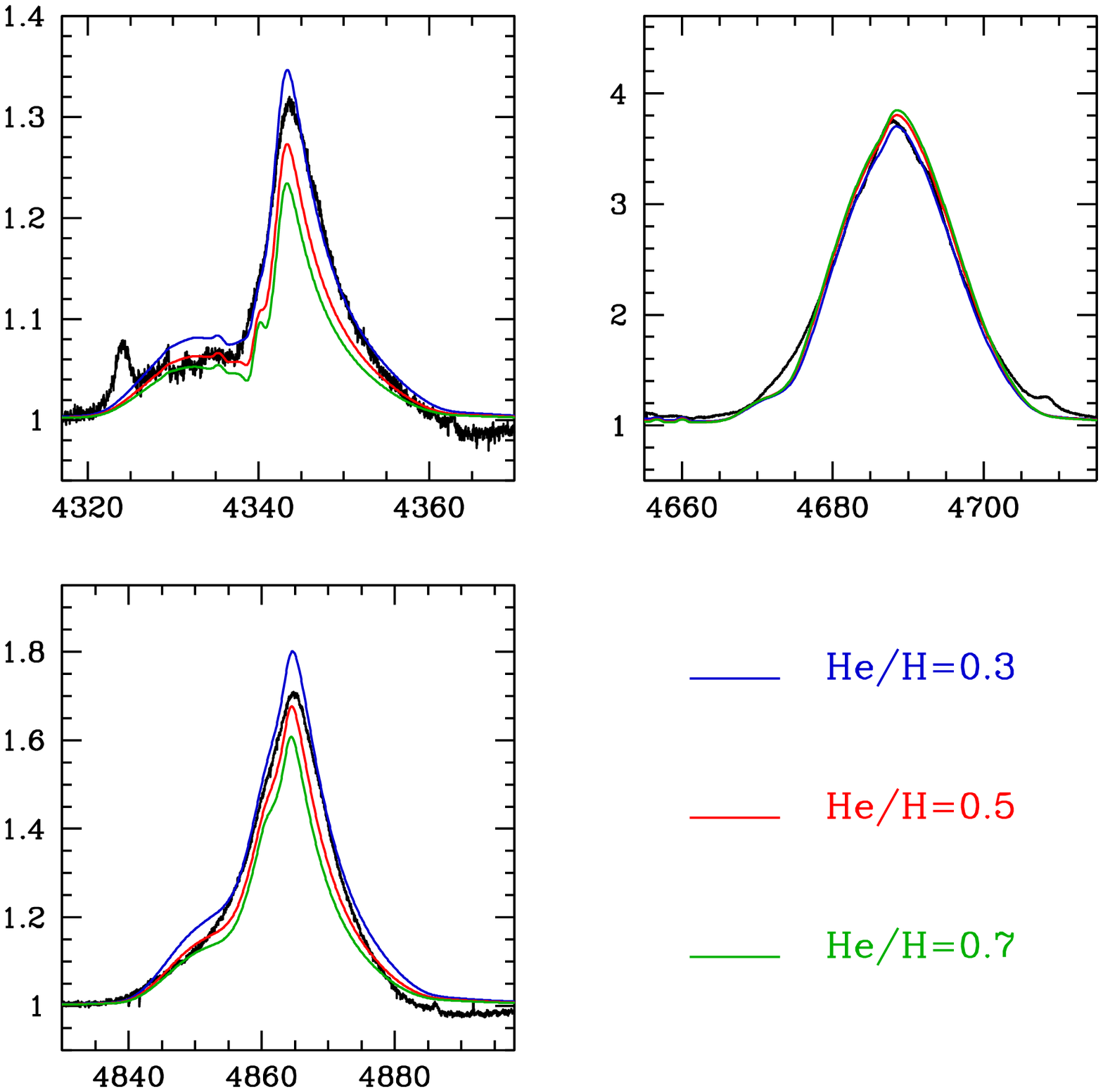}
\caption{Comparison between the observed spectrum of WR~10 (black line) and CMFGEN models with different He/H ratios. The lines are H$\beta$ (upper left), HeII~4686 (upper right) and H$\gamma$ (lower left). The mass loss rate of the models is adjusted so that HeII~4686 is always correctly reproduced.}
\label{effect_He_wr10}
\end{figure}

One of the key parameters we obtained is the hydrogen content of the stars. In Fig.\ \ref{effect_He_wr10} we illustrate the determination of the He/H ratio in the case of star WR~10. The models all reproduce correctly the HeII~4686 emission. For He/H=0.3, the hydrogen content is too large and the Balmer lines are too strong. Alternatively, for He/H=0.7, the hydrogen content is too small and the Balmer emission is too weak. Consequently, a value of He/H=0.5 is preferred. This also gives an estimate of the uncertainty on the He/H determination. We see that all WNh stars have a non negligible fraction of hydrogen in their atmosphere, while the two comparison WN4 stars are hydrogen free. 

The four Galactic stars were analyzed by \citet{hamann06}. We usually derive lower temperatures. The difference is of about 10000 K for WR10 and WR128, but amounts to 20000-30000 K for WR7 and WR18. It is difficult to increase significantly the temperature of our models for those objects since the \ion{Fe}{v} forest around 1300-1400 \AA\ becomes very weak, contrary to what is observed. On the other hand, \ion{N}{v} 4605--4620 is better reproduced at such high temperatures \citep[see also][]{bouret12}. 

The luminosities we derive for the Galactic stars are consistent with the results of \citet{hamann06}. Our mass loss rate determination are also in good agreement, with differences smaller than 0.2 dex. \citet{hamann06} did not find any evidence for the presence of hydrogen in WR~7 and WR~18, consistent with our determinations. For WR~10 and WR~128, they quote a hydrogen mass fraction of 0.25 and 0.16 respectively. This is again in good agreement with our results.

\begin{table*}[]
\begin{center}
\caption{Parameters of the target stars determined from spectroscopic analysis. } \label{tab_res}
\begin{tabular}{llrrrrrrrrrrrrr}
\hline
Star & ST               &  \teff & T$_{*}$ & \lL\ & R$_{*}$ & log(\mdot) & \vinf  & f & X(H) & He/H & X(C)            & X(N)            \\ 
     &                  &   [kK] & [kK]   &      & [\rsun] &            & [\kms] &   &      &      & [$\times 10^{-5}$] & [$\times 10^{-3}$]  \\
\hline
 Galaxy \\
\hline
WR7   & WN4             & 60.0   & 80.8   & 5.40  & 2.57   & -4.80     & 1600    & 0.1 & $<$0.01        & $>$10         & 14.7$\pm$5.0   & 11.0$\pm$5.0 \\
WR10  & WN5h            & 53.5   & 55.2   & 5.45  & 5.79   & -5.40/-5.45 & 1400  & 0.1 & 0.33$\pm$0.11  & 0.50$\pm$0.20 & 7.9$\pm$3.0    & 14.0$\pm$5.0 \\
WR18  & WN4             & 56.0   & 74.1   & 5.30  & 2.73   & -4.60     & 2200    & 0.3 & $<$0.01        & $>$10         & 17.7$\pm$5.0   & 10.3$\pm$4.0 \\
WR128 & WN4(h)          & 57.0   & 59.9   & 5.50  & 5.43   & -5.30     & 1800    & 0.1 & 0.26$\pm$0.07  & 0.70$\pm$0.20 & 6.2$\pm$1.6    & 11.0$\pm$3.0  \\
\hline
 LMC & \\
\hline
Bat 18 & WN3h           & 60.0   & 72.8   & 5.50  & 3.54   & -5.02     & 1800    & 0.3 & 0.25$\pm$0.10  & 0.80$\pm$0.20 & 5.7$\pm$2.0    & 6.6$\pm$2.1  \\
Bat 63 & WN4ha          & 58.5   & 68.9   & 5.45  & 3.73   & -5.45     & 2000    & 0.1 & 0.40$\pm$0.20  & 0.35$\pm$0.15 & 20.0$\pm$15.0  & 2.3$\pm$1.6  \\
\hline
\end{tabular}
\end{center}
\end{table*}

%%%%%%%%%%%%%%%%%%%%%%%%%%%%%%%%%%%%%%%%%%%%
\subsection{Evolutionary status}

\begin{figure*}[]
     \centering
\subfigure[]{ \includegraphics[width=.4\textwidth]{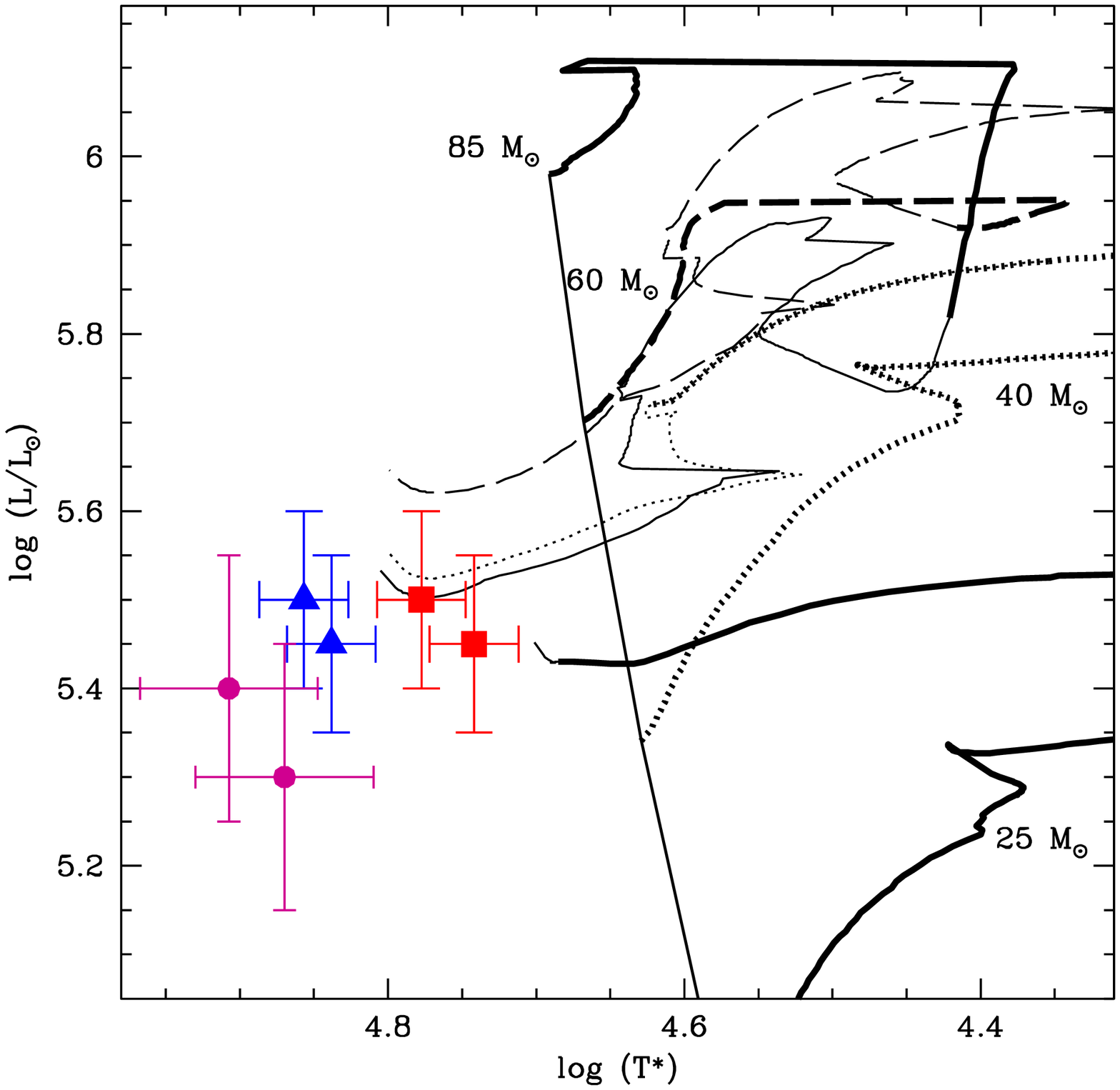}}
    \hspace{0.1cm}
\subfigure[]{\includegraphics[width=.4\textwidth]{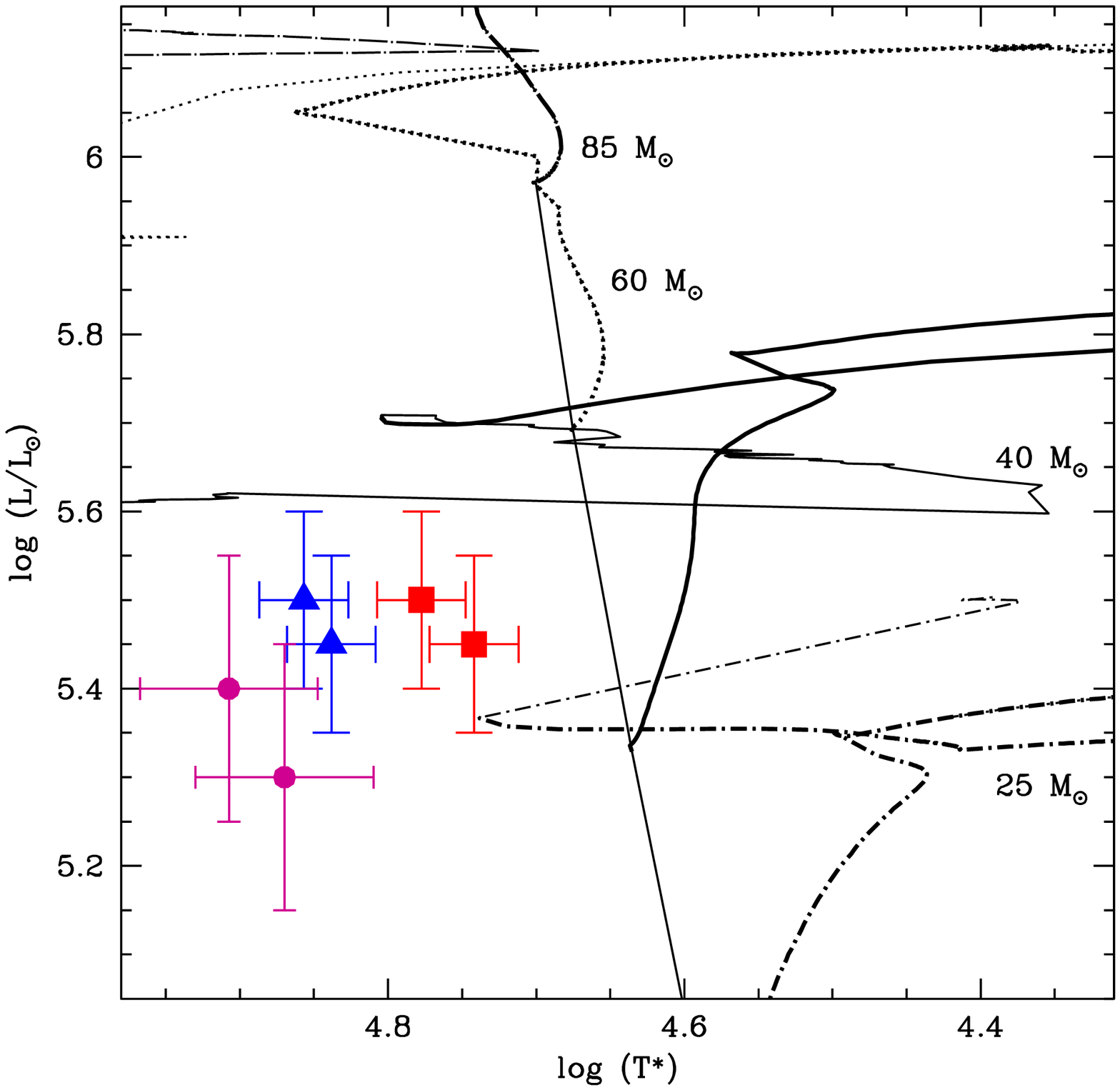}}
     \caption{HR diagram with the position of the sample WN stars shown by triangles (for LMC targets) and squares (Galactic targets). Circles correspond to the comparison Galactic WN4 stars. On the left panel, the evolutionary tracks are from \citet{mm05}, while on the right panel they are from \citet{ek12}. The bold part of the tracks corresponds to a hydrogen mass fraction larger than 0.1.}
     \label{hr_wnh}
\end{figure*}

\begin{figure}[]
\centering
\includegraphics[width=9cm]{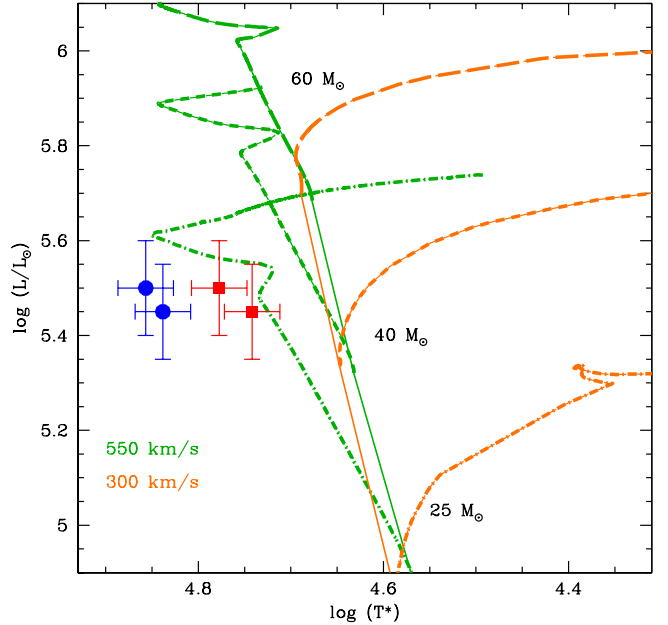}
\caption{HR diagram with the evolutionary tracks of \citet{brott11a} with an initial rotational velocity of 300 \kms\ (orange) and 550 \kms\ (green). The symbols have the same meaning as in Fig.\ \ref{hr_wnh}. The tracks all correspond to He/H$<$2.0 (equivalent of X(H)$\gtrsim$0.1). }
\label{hr_wnh_br_lmc}
\end{figure}

Fig.\ \ref{hr_wnh} shows the HR diagram with the position of the sample stars indicated by triangles (LMC) and squares (Galaxy). The evolutionary tracks are from Geneva and include rotation. Two sets of tracks are displayed: the ones by \citet{mm05} (left) and the ones by \citet{ek12} (right). The reference solar metallicity is Z=0.020 in the Meynet \& Maeder tracks, while it is Z=0.014 in the Ekstr$\ddot{o}$m et al. tracks. We show in bold the part of the evolutionary tracks for which the hydrogen mass fraction is higher than 0.1. This value is lower than the values we determine for all WNh stars of our sample. The Galactic stars cannot be reproduced by the 2005 tracks. Their position in the HR diagram is reached only by hydrogen-free tracks. With the more recent tracks, the stars are at the limit of the region reached by $\sim$25 \msun\ track still containing a fraction of hydrogen. They can be only marginally represented by those tracks. 
LMC tracks with normal (300 \kms) and fast (550 \kms) rotation are used to build the HR diagram shown in Fig.\ \ref{hr_wnh_br_lmc}. The two LMC stars (blue triangles) are clearly too hot to be reproduced by standard evolutionary tracks (orange lines) which never come back to the blue part of the HR diagram with a significant amount of hydrogen at their surface, contrary to what Bat~18 and Bat~63 show. 
From those comparisons, one can conclude that the LMC WNh stars, and most likely the Galactic ones too, cannot be explained by standard evolutionary tracks. This is similar to what we obtained for the SMC stars \citep{wnh09}. 

In this study, we concluded that quasi-chemically homogeneous evolution could be a way to explain the properties of the SMC early-type WNh stars. Such evolution takes place when the mixing timescale becomes shorter than the nuclear timescale. This is possible if the stars rotate initially very fast and do not lose too much angular momentum during their evolution. The Geneva tracks used so far exist only for an initial rotational velocity of 0 or 300 \kms. In Fig.\ \ref{hr_wnh_br_lmc}, we show the HR diagram built with the evolutionary tracks of \citet{brott11a} with a LMC composition. Those tracks have been computed for a variety of rotational velocities. We have plotted the tracks with initial velocities of 300 and 550 \kms. The latter tracks evolve directly to the blue part of the HR diagram from the main sequence. This is very different compared to the 300 \kms\ tracks. The positions of the stars is well reproduced by the fast rotating tracks for initial mass close to 25 \msun. In addition, they have X(H)$>$0.1, consistent with our spectroscopic analysis. We can thus conclude that the sample stars (at least those from the LMC) are consistent with a quasi-chemically homogeneous evolution. \citet{besten11} studied the WN5h LMC star VFTS682 and reached the same conclusion from its position in the HR diagram (they did not constrain the surface abundances).

\citet{brott11a} have shown that they could produce this type of blueward evolution only at metallicities below that of the LMC. Their fast rotating solar metallicity tracks behave more classically, with a redward extension from the main sequence. But if we look at Fig.\ \ref{hr_wnh}, we see that the recent Geneva tracks at solar metallicity tend to evolve towards the blue quite rapidly after the main sequence. The 2005 Geneva models used Z=0.020 while the 2012 tracks use the revised solar abundances, corresponding to Z=0.014. This might favour chemically homogeneous evolution at high rotation. And this leaves room for a pure homogeneous evolution for higher initial velocities even at solar metallicity.

\begin{figure}[]
\centering
\includegraphics[width=9cm]{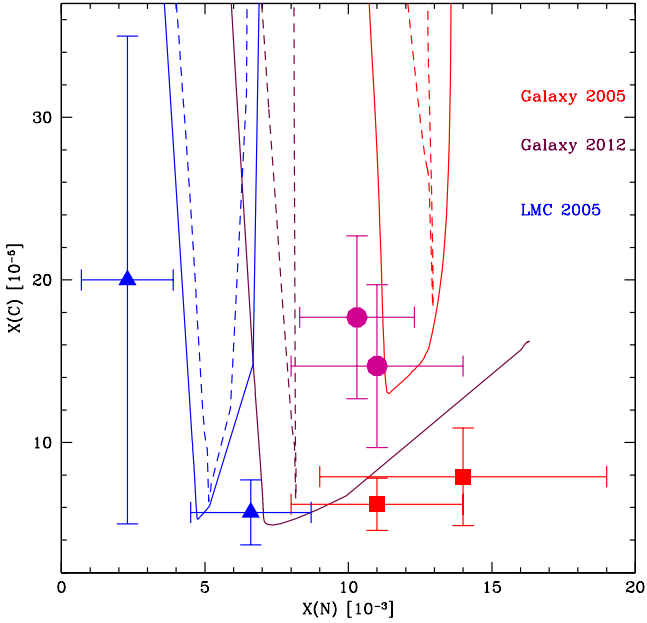}
\caption{Carbon mass fraction as a function of nitrogen mass fraction for the sample stars. The red squares (blue triangles) are the Galactic (LMC) objects. The Galactic comparison stars are shown by the purple circles. Evolutionary tracks from \citet{mm05} at solar (LMC) metallicity are shown by red (blue) lines. The solar metallicity tracks of \citet{ek12} are shown by the dark red lines. Solid and dashed lines correspond respectively to 25 (30 at LMC metallicity) and 60 \msun\ stars. }
\label{xc_xn}
\end{figure}

In Fig.\ \ref{xc_xn} we show the carbon mass fraction as a function of nitrogen mass fraction for our target stars. The rotating tracks from \citet{mm05} and \citet{ek12} are overplotted. The WNh stars are located in the part of the diagram corresponding to the minimum of carbon content and maximum of nitrogen content (for their respective metallicity). This is typical of stars showing products of the CN cycle at their surface. This is a very strong indication that the WNh stars are core H-burning objects. Indeed, if the stars evolve quasi chemically homogeneously and if they show CN equilibrium surface abundances, this means that their core CN abundances are also at equilibrium. Thus, the stars are still in the core H-burning phase. If they were more evolved, the central CN abundances would be no longer those of CN equilibrium.

 The properties of the sample early-type WNh stars are similar to the SMC WNh stars studied by \citet{wnh09}. They are relatively unevolved objects. WR~7 and WR~18, the comparison Galactic stars, are as nitrogen rich as the two Galactic WNh stars, but show a significantly larger carbon content. This is a convincing evidence that they are in a more advanced state of evolution since according to the theoretical prediction, after reaching a minimum, the carbon mass fraction starts to increase due to the onset of core He-burning. These Galactic WN stars certainly follow normal evolution and have experienced a classical redward evolution in the HR diagram before coming back to the left of the ZAMS, after having burnt all of their hydrogen. The surface composition of WN stars is thus a key diagnostic of their type of evolution.

%%%%%%%%%%%%%%%%%%%%%%%%%%%%%%%%%%%%%%%%%%%%%%%%%%%%%%%%%%%%%%%%%%%%%%%%%%%%%%%%%%%%%%%%%%%%%%%%%%%%%%%%%%%%%%%%%%%%%%%%%%%%%%%
%%%%%%%%%%%%%%%%%%%%%%%%%%%%%%%%%%%%%%%%%%%%%%%%%%%%%%%%%%%%%%%%%%%%%%%%%%%%%%%%%%%%%%%%%%%%%%%%%%%%%%%%%%%%%%%%%%%%%%%%%%%%%%%
\section{Discussion}
\label{s_disc}

Our results show that CHE takes place in galaxies with average metallicities higher than the SMC. Since metallicity gradients exist within galaxies, this does not necessarily mean that our sample stars are located in region with large metal content. Although we have assumed that the LMC stars have a metal content 0.5 times the solar value, we have tried to obtain better estimates for the Galactic objects.
We have used the position of the Galactic stars
in the disk to estimate their metallicity from known O/H and Fe/H gradients established from Planetary Nebulae, HII regions and cepheids.
Planetary Nebulae and HII regions
are two types of objects which are commonly used to determine the abundance gradient in the Milky Way disk.
The oxygen abundance, which is easily observed in such emission-line objects, is expected to give the interstellar matter abundance at the time of the formation of the stars.
Cepheids provide the iron content as a function of galactocentric radius \citep{lemasle07}.
The results are given in Table \ref{tab_Z}.
For HII regions \citep{balser11}, we provide two values: one from the azimuthally
averaged gradient (Zhii) and one from the relation established in the azimuthal range corresponding to the star's galactic longitude (Zhiib). For Planetary nebulae, we use the relation of \citet{henry10}.  
The O/H and Fe/H determinations give consistent results. From Table \ref{tab_Z}, we clearly see that WR~128 has a near solar composition. WR~10 is located in a slightly less enriched medium, with Z$\sim$0.60-0.65. Together with the LMC targets assumed to have a metallicity of 0.5 the solar value, we conclude that the WNh stars following quasi-chemically homogeneous evolution are found in the interval 0.5-1.0 Z$_{\odot}$.

We have not determined the metal content of each star directly from its spectrum. The \ion{Fe}{v} and \ion{Fe}{vi} lines in the UV range could be used in principle. However, only the \ion{Fe}{v} forest around 1350-1400 \AA\ is not dominated by strong emission lines. With only a single ionization state available, the determination of the Fe content is difficult since temperature uncertainties (of the order 3000 K) strongly affect the estimate of metallicity. Thus, we think metallicity estimates based on galactic gradients are a better than any direct diagnostic we have.

\begin{table}[]
\begin{center}
\caption{Metallicity estimates} \label{tab_Z}
\begin{tabular}{lrrrrrrrrrr}
\hline
Star  &     d   &  R(kpc) & Zpn    &  Zhii  & Zhiib  & Zcp\\ 
      &  [kpc]  &  [kpc]  &        &        &        & \\
\hline
WR128 &  4.30   &  7.03   & 1.05  & 0.97  & 1.07  & 1.13 \\
WR18  &  3.30   &  8.37   & 0.88  & 0.85  & 0.81  & 0.95 \\
WR10  &  4.60   &  11.19  & 0.60  & 0.64  & 0.65  & 0.66 \\
WR7   &  4.80   &  11.30  & 0.52  & 0.57  & 0.60  & 0.65 \\
\hline
\end{tabular}
\tablefoot{Z represents the metallicity. Zpn is obtained from from Planetary Nebulae; Zhii and Zhiib from HII regions (see text for discussion); Zcp from cepheids. It is expressed in oxygen abundance assuming 12+log(O/H) = 8.66 for the Sun \citep{ga07} for planetary nebulae and HII regions, and in iron abundance for cepheids.}
\end{center}
\end{table}

Our analysis provides good evidence for the occurrence of quasi-chemically homogeneous evolution at solar metallicity. The current evolutionary tracks at solar metallicity do not predict such an evolution \citep{brott11a,ek12}. Fig.\ 5 of \citet{brott11a} clearly shows that a blueward evolution for fast rotating stars is possible only at metallicities below 0.5 Z$_{\odot}$. This is usually interpreted as a wind effect. At higher metallicity, mass loss rates of massive stars are higher \citep{mokiem07} and thus the removal of angular momentum is easier. Stars brake faster and are never fully mixed, the necessary condition to have a homogeneous evolution. However, this interpretation assumes that the core and the envelope are strongly coupled, so that the reduction of the surface angular momentum by stellar wind is mirrored in the core angular momentum. If the coupling is weak or at least moderate, the effect of winds is reduced \citep[see discussion in][]{georgy12}. The models of \citet{brott11a} include the transport of angular momentum due to the presence of a magnetic field. This strengthens the core-envelope coupling, at least partially explaining that the occurrence of quasi-chemically homogeneous evolution is correlated with metallicity and wind strength. \citet{petrovic05b} also studied the angular momentum evolution of single and binary stars and concluded that the presence of magnetic field favoured a slowly rotating core.

The solar metallicity (Z=0.014) models of \citet{ek12} seen in Fig.\ \ref{hr_wnh} (right) show a clear deviation to the blue slightly after the zero-age main sequence and above 40 \msun. These models do not include magnetic coupling. Qualitatively, they indicate that quasi-chemically homogeneous evolution is not excluded, since they are somewhat intermediate between fully chemically mixed models and models with normal mixing. Given the uncertainties in the mass loss rates and the physics of angular momentum transport, it is thus not excluded that evolutionary models can reproduce quasi-chemically homogeneous evolution.

According to the present results, to reproduce the properties of the early WNh stars in the Galaxy, the following conditions are required: 1) a very efficient mixing of chemical species; 2) a fast initial rotation; 3) a moderate or weak angular momentum coupling between the stellar core and the envelope. In such a way, the effect of stellar winds are reduced to the outer layers. This can explain that our WNh stars do not seem to be extremely fast rotators.

The determination of \vsini\ for our objects is very difficult since most lines are formed throughout the wind and are broadened by its expansion. Polarimetry sometimes indicates that Wolf-Rayet stars have a flattened shape probably due to rotation. Some of these objects may be identified as precursors of LGRBs \citep{vink11,gv12}. However, even without polarimetric information, we can determine orders of magnitudes for the projected rotational velocity and clearly state that \vsini\ values larger than 400 \kms are safely excluded. Above such a value the details of some emission line profiles (such as \ion{C}{iv} 5802-5812) are smoothed out. The surface projected velocities of our sample stars are thus of a few tens to a couple of hundreds of \kms at most. Hence, a moderately low surface rotational velocity is not necessarily in contradiction with a quasi-chemically homogeneous evolution. 

If the angular momentum coupling between the core and the envelope is only moderate, a star evolving homogeneously may form a collapsar \citep{ww93} and a long-soft gamma-ray burst. Recent observations reveal that such events are not restricted to low metallicity environments. \citet{levesque10} report the occurrence of a LGRB in a galaxy with 12+log(O/H)=9.0, larger than the solar value. Similarly, \citet{mannucci11} show that the mass--luminosity relation of LGRB host does not stop at solar metallicity: LGRB hosts are simply more massive for a given metallicity. Given the effects of stellar winds described above, it is likely that lower Z favours CHE because braking is less efficient (even in the case of weak coupling, there is still an effect). But this type of evolution can also take place in solar metallicity environments. The most recent compilation of Wolf-Rayet stars by P. Crowther \footnote{http://pa crowther.staff.chef.ac.uk/WRcat/} contains 433 objects, among which only seven are of spectral type WN3h, WN4h or WN5h. If one assumes that such stars have similar properties to our sample stars, one may expect only 1-2\% of Wolf-Rayet stars to be massive stars evolving quasi homogeneously. This is most likely an upper limit. Indeed, WN3-5h stars being core-H burning objects, their number should be compared to the number of normal main sequence stars, which is larger than the number of Wolf-Rayet stars. In the LMC, the catalog of \citet{brey99} lists 10 stars with spectral type WN3-5h among 134 objects, corresponding to a fraction of 7\% of Wolf-Rayet stars being hydrogen-rich early-type WN objects. Even if a complete census of the Galactic Wolf-Rayet is still missing, it seems that the fraction of WN3-5h is larger in the LMC compared to the Galaxy. Assuming that all WN3-5h stars experience CHE, this is consistent with the theoretical prediction that CHE takes place more likely -- but not exclusively -- at low metallicity \citep{brott11a}. 

Fully mixed massive stars should be observed in the O star phase when they are close to the zero-age main sequence. There are several indications that this is the case. \citet{bouret03} found that the hot O2III(f*) star MPG~355 in the SMC cluster NGC~346 is much younger than the bulk of O stars (1 versus 3 Myr). They argue that it evolves homogeneously, which is consistent with its relatively high nitrogen enhancement triggered by fast rotation. Consequently, they follow isochrones much more ``vertical'' in the HR diagram and their age cannot be derived by standard isochrones. Similar conclusions are reached by \citet{walborn04} concerning ON2III stars. They are located close to the ZAMS in the HR diagram and feature a large N/C ratio. \citet{mokiem07} studied a sample of LMC stars and reported a correlation between high helium enrichment and presence of a mass discrepancy (difference between masses derived from spectroscopy and from the evolutionary tracks). A natural explanation is that the use of normally rotating tracks for stars actually following homogeneous evolution will lead to a strong overestimate of the evolutionary mass. They thus conclude that some of the LMC O stars could be fully mixed. They also note that three of their O2 stars are located blueward of the main sequence. The indirect evidence for quasi-chemically homogeneous evolution in the O star phase are rather compelling, at least in the Magellanic Clouds. Similar results for Galactic O stars do not (yet) exist. 

An alternative explanation for the position of the WNh stars in the HR diagram is binarity. Evolutionary calculations of massive binary systems indicate that as soon as the two components interact through mass transfer, the evolutionary tracks are severely modified compared to single star tracks \citep{wellstein01,petrovic05}: the accretor can turn back to the blue part of the HRD. Recent surveys of young star clusters indicate that a significant fraction -- ranging from 30 to 70\% -- of massive stars have a companion \citep{sana12,sana13}. Mass transfer (and even merging) can occur during the early phases of evolution in short-period systems \citep{demink12}. In such a case, the accreting star receives mass and angular momentum. It can appear bluer than the bulk of the surrounding population, just as blue stragglers in globular clusters. This could explain the position of our sample WNh stars in the HR diagram. We have seen that the surface composition of the WNh stars was typical of CN equilibrium. In the binary scenario, the surface appearance of the accreting star depends strongly on the phase at which mass transfer occurs. If it is during the expansion of the primary, at the end of central hydrogen burning, the material dropped onto the secondary can show the signatures of CN processing. But depending on the initial period of the binary system, mass transfer can occur either before or much after hydrogen core exhaustion. In that case, no special surface abundance pattern is expected. It appears unlikely that binarity can explain the properties of \textit{all} the early-type WNh stars we have studied. \citet{foel03} studied the binary status of Bat~18 and Bat~63 and found no evidence for radial velocity nor photometric variations, consistent with them being single stars. No clear evidence for binarity exists for WR~10 and WR~128 either. We thus conclude that binarity could in principle explain the properties of some early-type WNh stars, but the ones we studied here are better accounted for by CHE.  

All in all, there are now several robust indications that quasi-chemically homogeneous evolution happens at different metallicities, up to solar. An understanding of the conditions under which it appears is necessary since the current evolutionary models do not predict it at high metallicity. Beyond the understanding of a somewhat limited number of objects, their study is of importance to better constrain the mixing mechanism in massive stars. With better predictions, it will be easier to interpret data on the occurrence of LGRBs in various environments.

%%%%%%%%%%%%%%%%%%%%%%%%%%%%%%%%%%%%%%%%%%%%%%%%%%%%%%%%%%%%%%%%%%%%%%%%%%%%%%%%%%%%%%%%%%%%%%%%%%%%%%%%%%%%%%%%%%%%%%%%%%%%%%%
%%%%%%%%%%%%%%%%%%%%%%%%%%%%%%%%%%%%%%%%%%%%%%%%%%%%%%%%%%%%%%%%%%%%%%%%%%%%%%%%%%%%%%%%%%%%%%%%%%%%%%%%%%%%%%%%%%%%%%%%%%%%%%%
\section{Conclusion}
\label{s_conc}

We have presented the analysis of early-type H-rich WN stars in the LMC and the Galaxy. The stellar and wind parameters have been determined through spectroscopic analysis in the UV and optical range. Atmosphere models and synthetic spectra computed with the code CMFGEN have been used. The main results are:

\begin{itemize}

\item[$\bullet$] the early-type WNh stars are hot objects located on the blue part of the zero age main sequence.

\item[$\bullet$] in spite of their Wolf-Rayet classification, they still contain a significant amount of hydrogen (X(H) $>$ 0.2). In addition, their C and N composition is typical of the CN equilibrium. Altogether, this indicates that the WN3-5h stars are core-hydrogen burning objects. Comparison to H-free WN4 stars indicates clear differences in the evolutionary state of both types of objects (WNh versus WN).

\item[$\bullet$] classical evolutionary tracks cannot reproduce both the position in the HR diagram and the chemical composition of these objects. Fully mixed models evolving blueward from the zero-age main sequence are consistent with such properties. WN3-5h stars most likely follow quasi-chemically homogeneous evolution

\item[$\bullet$] estimates of the metallicity in the direct environment of the target stars reveals 0.5 $<$ Z $<$ 1.0, indicating that quasi-chemically homogeneous evolution happens at solar metallicity. This favours a relatively moderate coupling between the core and the envelope for angular momentum transport, since the effect of stronger stellar winds at higher metallicity does not prevent quasi-chemically homogeneous evolution, contrary to the current predictions of evolutionary models. This is also consistent with the ``not extreme'' rotational velocity of the WN3-5h stars.

\item[$\bullet$] the finding that quasi-chemically homogeneous evolution takes place in solar metallicity environments and the discovery of LGRBs in metal-rich galaxies support the scenario according to which the progenitors of LGRBs result from quasi-chemically homogeneous evolution.

\end{itemize}

The total number of WN3-5h stars in the Galaxy is relatively small (1-2\% of all Wolf-Rayet stars). If all WN3-5h stars follow quasi-chemically homogeneous evolution, this shows that this type of evolution is rare. It is however crucial to understand it, not only in the context of LGBR production, but also to better constrain the physical processes in stellar interior, especially regarding mixing and angular momentum transport. Future theoretical studies should help improve our knowledge of such effects.

%%%%%%%%%%%%%%%%%%%%%%%%%%%%%%%%%%%%%%%%%%%%%%%%%%%%%%%%%%%%%%%%%%%%%%%%%%%%%%%%%%%%%%%%%%%%%%%%%%%%%%%%%%%%%%%%%%%%%%%%%%%%
%%%%%%%%%%%%%%%%%%%%%%%%%%%%%%%%%%%%%%%%%%%%%%%%%%%%%%%%%%%%%%%%%%%%%%%%%%%%%%%%%%%%%%%%%%%%%%%%%%%%%%%%%%%%%%%%%%%%%%%%%%%%
\begin{acknowledgements}
We thank an anonymous referee for comments which helped to improve the clarity of our statements. We thank John Hillier for making his code CMFGEN available and for constant help with it. FM acknowledges support from the ``Agence Nationale de la Recherche (ANR)''. We thank C\'edric Foellmi for providing us with the optical spectra of stars Bat~18 and Bat~63. 
\end{acknowledgements}

%%%%%%%%%%%%%%%%%%%%%%%%%%%%%%%%%%%%%%%%%%%%%%%%%%%%%%%%%%%%%%%%%%%%%%%%%%%%%%%%%%%%%%%%%%%%%%%%%%%%%%%%%%%%%%%%%%%%%%%%%%%%
%%%%%%%%%%%%%%%%%%%%%%%%%%%%%%%%%%%%%%%%%%%%%%%%%%%%%%%%%%%%%%%%%%%%%%%%%%%%%%%%%%%%%%%%%%%%%%%%%%%%%%%%%%%%%%%%%%%%%%%%%%%%
\bibliographystyle{aa}
\bibliography{wr_hom}

%%%%%%%%%%%%%%%%%%%%%%%%%%%%%%%%%%%%%%%%%%%%%%%%%%%%%%%%%%%%%%%%%%%%%%%%%%%%%%%%%%%%%%%%%%%%%%%%%%%%%%%%%%%%%%%%%%%%%%%%%%%%
%%%%%%%%%%%%%%%%%%%%%%%%%%%%%%%%%%%%%%%%%%%%%%%%%%%%%%%%%%%%%%%%%%%%%%%%%%%%%%%%%%%%%%%%%%%%%%%%%%%%%%%%%%%%%%%%%%%%%%%%%%%%
%\Online
%\begin{appendix}
%\end{appendix}

%%%%%%%%%%%%%%%%%%%%%%%%%%%%%%%%%%%%%%%%%%%%%%%%%%%%%%%%%%%%%%%%%%%%%%%%%%%%%%%%%%%%%%%%%%%%%%%%%%%%%%%%%%%%%%%%%%%%%%%%%%%%
%%%%%%%%%%%%%%%%%%%%%%%%%%%%%%%%%%%%%%%%%%%%%%%%%%%%%%%%%%%%%%%%%%%%%%%%%%%%%%%%%%%%%%%%%%%%%%%%%%%%%%%%%%%%%%%%%%%%%%%%%%%%
\end{document}